\documentclass{elsart}
\usepackage{graphicx}
\usepackage{amssymb}
\usepackage{epsfig}
\usepackage{subfigure}
\usepackage{color}
\usepackage[english]{babel}
\usepackage{latexsym}
\usepackage{amsfonts}
\usepackage{amsmath}
\usepackage{multirow}
\usepackage{float}

\def\Journal#1#2#3#4{{#1} {\bf #2} (#3) #4}

\def\PRL{Phys. Rev. Lett.}
\def\PRC{Phys. Rev. C}

\def\be{\begin{equation}}
\def\ee{\end{equation}}

\begin{document}
\begin{frontmatter}

\title{Production of light nuclei, hypernuclei and their antiparticles in 
relativistic nuclear collisions}

\author[gsi]{A.~Andronic},
\author[gsi,emmi,tud,fias]{P.~Braun-Munzinger},
\author[hei]{J.~Stachel},
\author[gsi,fias]{H.~St\"ocker}

\address[gsi]{GSI Helmholtzzentrum f\"ur Schwerionenforschung,
D-64291 Darmstadt, Germany}
\address[emmi]{EMMI, GSI, D-64291 Darmstadt, Germany}
\address[tud]{Technical University Darmstadt, D-64289 Darmstadt, Germany}
\address[fias]{Frankfurt Institute for Advanced Studies, J.W. Goethe University, D-60438 Frankfurt, Germany}
\address[hei]{Physikalisches Institut der Universit\"at Heidelberg,
D-69120 Heidelberg, Germany}

\begin{abstract}
  We present, using the statistical model, an analysis of the production of
  light nuclei, hypernuclei and their antiparticles 
  in central collisions of heavy nuclei. Based on these studies 
  we provide predictions for the production yields of multiply-strange 
  light nuclei.
\end{abstract}

\end{frontmatter}

\section{Introduction}
One of the major goals of ultrarelativistic nuclear collision studies is to
obtain information on the QCD phase diagram \cite{pbm_wambach}.  
Currently, one of the most direct
approaches is the investigation of hadron production.  Hadron yields measured in
central heavy ion collisions from AGS up to RHIC energies can be described very
well \cite{agssps,satz,heppe,cley,rhic,nu,beca2,rapp,becgaz,aa05}
within a hadro-chemical equilibrium model.  In our approach
\cite{agssps,heppe,rhic,aa05,aa08,aa09} the only parameters are the chemical 
freeze-out temperature $T$ and the baryo-chemical potential $\mu_b$ (and 
the fireball volume $V$, in case yields rather than ratios of yields are 
fitted).
Other approaches  \cite{beca1,beca2,becgaz,man08,letessier05} employ 
(several) other, non-thermal, parameters. For a review see \cite{review}.

The main result of these investigations was that the extracted temperature
values rise rather sharply from low energies on towards 
$\sqrt{s_{NN}}\simeq$10 GeV and reach afterwards constant values near 
$T$=160 MeV, while the baryochemical potential  decreases smoothly as a 
function of energy.
This limiting temperature \cite{hagedorn85} behavior suggests a connection to 
the phase boundary and it was, indeed, argued \cite{wetterich} that the 
quark-hadron phase transition drives the equilibration dynamically, at least 
for SPS energies and above. 
For the lower energies, the quarkyonic state of matter \cite{mclerran_pisarski} 
could complement this picture by providing a new phase boundary at large
$\mu_b$ values. The conjecture of the tricritical point \cite{tri} was put 
forward in this context.

The importance of measurements at very high energies  to obtain information on 
the existence of a limiting temperature of excited hadronic matter produced in 
nuclear collisions was pointed out early 
\cite{siem_kap,Stoecker:1981,Hahn:1986pw,Hahn:1986mb}
based on analysis of particle spectra at the Bevalac (see also the 
review \cite {Stoecker:1986ci}), from pions to heavier complex nuclei.

At first glance, it may seem inappropriate to use the chemical freeze-out
concept for light nuclei, as their binding energies are a few MeV, 
much less than the chemical freeze-out temperatures of 100-170 MeV.
We note, however, that the relative yield of particles composed of nucleons
is determined by the entropy per baryon, which is fixed at chemical freeze-out.
This has been first recognized already 30 years back \cite{siem_kap} and 
was subsequently further substantiated in \cite{Hahn:1986mb}, constituting
the basis of thermal analyses of yields of light nuclei \cite{pbm95,pbm01}.
It is entropy conservation, and not the difference between the binding energy and
temperature of the system, which governs the production yields in this case 
After chemical freeze-out, entropy is conserved.

It was also noted then that the yields obtained within the thermal model
are in close agreement to those from  coalescence models 
\cite{pbm95,baltz_dover}.
The thermal model studies were already at that time extended to nuclei carrying 
strangeness (hyperons in replacement of nucleons) and even hypothetical objects 
with roughly equal number of up, down and strange quarks (strangelets).
At the same time, a vigorous line of theoretical investigations on the existence 
of multi-strange hypernuclei, or MEMOs \cite{schaffner_gal_dover,gal_dover,gerland,memo}
was established.

Recently, the first measurement of the lightest (anti)hypernucleus, 
(anti)hyper-tritium, in high-energy nucleus-nucleus collisions was achieved by 
the STAR experiment at the RHIC \cite{star_hyp}.
This measurement opens up a very interesting new regime for tests of particle 
production at chemical equilibrium. At relatively low beam energies, where the 
baryo-chemical potential and, hence, the baryon density is maximum (FAIR energy
regime) objects with a large number of baryons and moderate strangeness may be
abundantly produced \cite{memo}. At RHIC and LHC energies production of
objects with moderate (anti)baryon number and large strangeness content may be expected.
In this paper we investigate and predict within the thermal model the production 
yields of heavy baryons and anti-baryons and in particular of hypernuclei and
their antiparticles  and
confront these calculations with all presently available data ranging from AGS
to RHIC energies.

\section{Preliminaries}

\begin{figure}[hbt]
\centering\includegraphics[width=.58\textwidth]{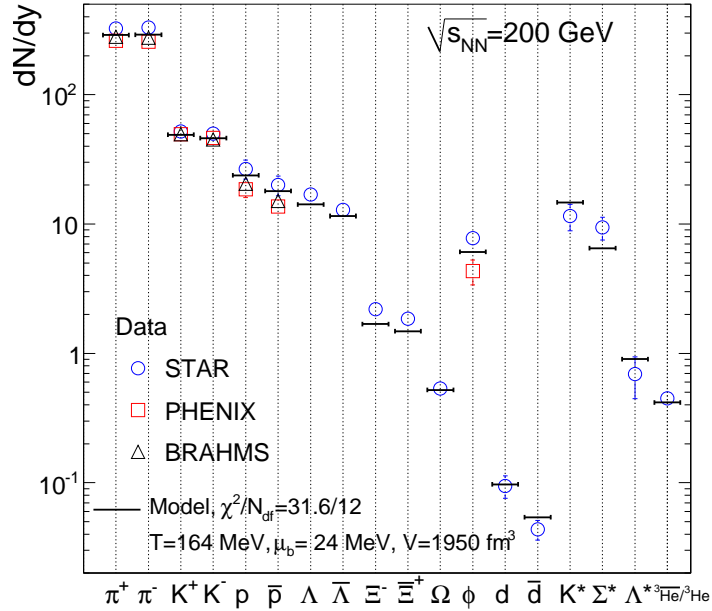}
\caption{Hadron yields in comparison with the thermal model fit of combined 
data (excluding $K^*$, $\Sigma^*$, $\Lambda^*$), for the
RHIC energy of $\sqrt{s_{NN}}$=200 GeV. The ratio $^{3}\bar{He}/^{3}He$, 
recently measured by the STAR experiment \cite{star_hyp} is included in the fit.}
\label{fig1}
\end{figure}

The measurement of the production yields of light nuclei (and anti-nuclei)
without strangeness in central nuclear collisions provides significant
constraints on thermal model parameters, in particular on the value of the
baryo-chemical potential $\mu_b$. This is most easily seen when one recognizes
that yield ratios such as $^{n}\bar{He}/^{n}He$ scale like $\exp{[-(2n\mu_b/T)]}$.
In Fig.~\ref{fig1} we show the updated thermal fit to the hadron yield data
measured at RHIC ($\sqrt{s_{NN}}$=200 GeV) including the newly-measured
\cite{star_hyp} yield ratio $^{3}\bar{He}/^{3}He$. Including this ratio
significantly narrows the range of possible $\mu_b$ values, while $T$ and $V$
of the new fit remain unchanged ($T$=164 MeV, $V$=1960 fm$^3$) compared to our
earlier fit \cite{aa08}.  Quantitatively, the new fit leads to
$\mu_b$=24$\pm$2 MeV, while without the ratio $^{3}\bar{He}/^{3}He$,
$\mu_b$=30$\pm$4 MeV \cite{aa08}. The quality of the present fit is similar to
that of the earlier one (which had $\chi^2$/dof=29.7/12). This result supports
previous findings at lower energies \cite{pbm95,pbm01,baltz_dover}. 
We stress that the agreement between the experimental value and the
calculated one for the ratio  $^{3}\bar{He}/^{3}He$ is a powerful argument 
that indeed entropy conservation governs the production of light nuclei.
If one were to use a temperature comparable to the binding energy per nucleon, 
that is $T$=5 MeV, the calculated ratio would be 3.1$\cdot$10$^{-13}$, while 
it is 0.415 for $T$=164 MeV, see Fig.~\ref{fig1}.

\begin{figure}[hbt]
\centering\includegraphics[width=.58\textwidth]{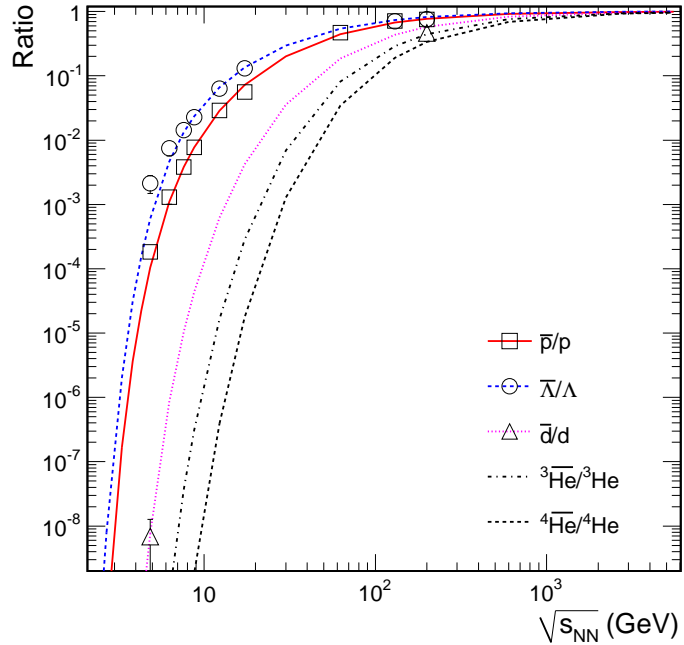}
\caption{Energy dependence of anti-baryon to baryon yield ratios. The lines
  are thermal model results as described in the text. The symbols represent
measured data.}
\label{fig2}
\end{figure}

In Fig.~\ref{fig2} we show that predictions using the thermal model can be
used to describe quantitatively the measured energy dependence of $\bar p/p$,
$\bar d/d$, and $\bar\Lambda/\Lambda$ yield ratios over a very wide energy range
The calculations, here and in the following, are performed using the 
parametrizations for $T$ and $\mu_b$ established in \cite{aa08} based on 
fits of midrapidity data in central collisions.
The penalty against anti-particles at lower energies, well described within 
the thermal approach, is also drastically exhibited in this figure. 
Note, in particular, the very good agreement between the model and the 
measurements at AGS \cite{e864d} for the $\bar d/d$ ratio, which extends the
the range of the model agreement over 8 orders or magnitude
(for the other measurements see references in \cite{aa05,aa08}).
It is thus natural to extend such analyses to light nuclei containing strangeness.
Note that, at the lowest energies, the canonical strangeness  suppression 
is important and is incorporated in our model as described in \cite{aa05}.
In the following we will therefore use the thermal model with the parameters as 
discussed to analyze the production of hypernuclei and their anti-particles and 
confront model predictions with the now available data.

\section{Production of Hypernuclei at RHIC Energy}

In Table~\ref{tab1} we show a comparison of the measured data \cite{star_hyp}
and model calculations for the best fit thermal parameters discussed
above. The yield ratio $^{3}_{\bar{\Lambda}}\bar{H}/^{3}_{\Lambda}H$ is as
well reproduced by the model as the ratio $^{3}\bar{He}/^{3}He$.  On the other
hand, the measured ratios of (anti)hyper-tritium to (anti)$^3He$ are larger
than predicted in the model by about two standard deviations, using the
statistical and systematic uncertainties quoted for the data.

\begin{table}[hbt]
\caption{Ratios at RHIC energy, $\sqrt{s_{NN}}$=200 GeV. 
The experimental values are from the STAR experiment 
\cite{star_hyp} and contain statistical ans systematic errors. The errors for
the model calculations correspond to the errors of the fit for the baryochemical 
potential, $\mu_b=24\pm2$ MeV.}
\label{tab1}
\begin{tabular}{lcc}
Ratio                & Experiment & Model \\ \hline
$^{3}\bar{He}/^{3}He$ & 0.45$\pm$0.02$\pm$0.04  & 0.42$\pm$0.03 \\ 
$^{3}_{\bar{\Lambda}}\bar{H}/^{3}_{\Lambda}H$ & 0.49$\pm$0.18$\pm$0.07  & 0.45$\pm$0.03 \\ 
$^{3}_{\Lambda}H/^{3}He$ & 0.82$\pm$0.16$\pm$0.12  & 0.35$\pm$0.003 \\ 
$^{3}_{\bar{\Lambda}}\bar{H}/^{3}\bar{He}$ & 0.89$\pm$0.28$\pm$0.13  & 0.37$\pm$0.003 \\ 
\end{tabular}
\end{table}

\begin{figure}[hbt]
\centering\includegraphics[width=.6\textwidth]{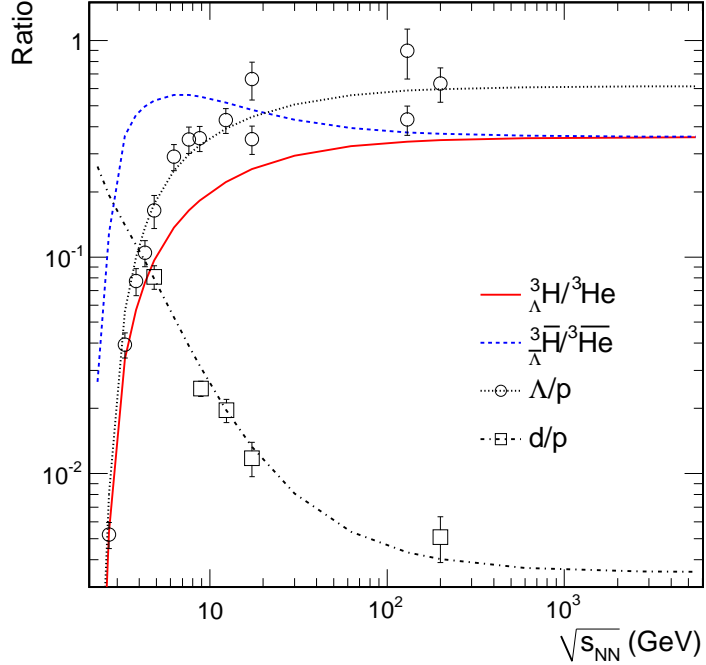}
\caption{Energy dependence of various baryon yield ratios. The lines are 
calculations, the symbols are experimental data.}
\label{fig3}
\end{figure}

To shed more light on the situation we turn now to the energy dependence of
(strange) baryon production.
In Fig.~\ref{fig3} we show the experimental energy dependence of the
$\Lambda/p$ and $d/p$ ratios and confront these data with our thermal model 
predictions. The degree of agreement between data and calculations 
is impressive. We also include in this figure thermal model predictions for 
the energy dependence of the ratio $^{3}_{\Lambda}{H}/^{3}{He}$ and 
$^{3}_{\bar{\Lambda}}\bar{H}/^{3}\bar{He}$.  
The broad maximum around $\sqrt{s_{NN}}\simeq$5 GeV for the ratio
$^{3}_{\bar{\Lambda}}\bar{H}/^{3}\bar{He}$ has the same origin as the
maximum in the $K^+/\pi^+$ ratio, namely it arises as a consequence of 
strangeness neutrality condition, imposed in our model, and a competition 
between rising $T$ and decreasing $\mu_b$ \cite{aa08}. We also note that a
slightly less prominent maximum is likely to survive even  if one relaxes 
the condition of strangeness neutrality, as demonstrated in \cite{memo}. 
At high energies the value for the two ratios approach each other, as expected 
for decreasing values of $\mu_b$ at a nearly constant temperature.

\begin{figure}[hbt]
\centering\includegraphics[width=.6\textwidth]{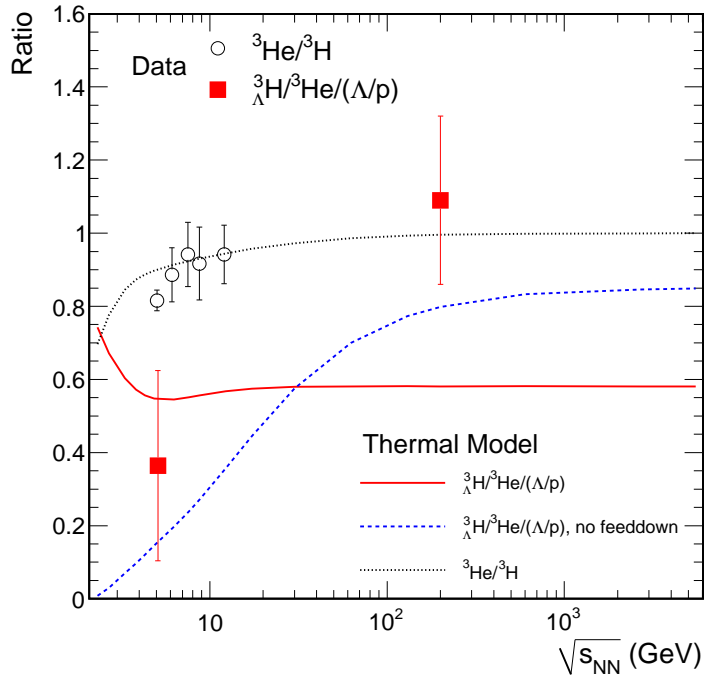}
\caption{Energy dependence of nuclei and hypernuclei production ratios. 
The data points are extracted from ref.~\cite{star_hyp}, the lines are 
our model calculations. Note the discrepancy between data and model for 
the double ratio involving hyper-tritium, where the continuous line represents 
the physical case, while the dashed line represents the case without the 
important contribution from feed-down from strong decays on the $\Lambda/p$
ratio (see text).}
\label{fig4}
\end{figure}

In Fig.~\ref{fig4} we show the measured energy dependence for the $^3He/^3H$
and the $^3_{\Lambda}H/(^3He (\Lambda/p))$ ratio. This double ratio was
suggested by the authors of \cite{star_hyp} in the expectation that dividing
out the strange to non-strange baryon yield should result in a value near
unity.  The data are compared to thermal model predictions. Note that there is
negligible feed-down from heavier states into states with baryon number 3.  
As expected, the measured energy dependence of the $^3He/^3H$ is well reproduced
by the model calculations. On the other hand, the discrepancy between thermal
model predictions and data for the $^3_{\Lambda}H/(^3He (\Lambda/p))$ ratio is
apparent (red line).  It is important to realize that the ratio $\Lambda/p$ is
significantly influenced by feed-down from strong decays of excited baryonic
states, leading to a value for the double ratio significantly below unity. 
The red line in Fig.~\ref{fig4} contains such feed-down, the blue dashed
line represents a calculation where the feeding is artificially left out. The
feed-down from strong decays increases the $\Lambda/p$ ratio and, hence,
reduces the overall ratio. The STAR collaboration actually measures a ratio
close to 1, above the thermal model prediction by twice the error quoted by
the experiment. Interestingly, results from the E864 collaboration \cite{e864hyp}
(as shown in ref. \cite{star_hyp}) at AGS energy are, albeit with large 
uncertainties, consistent with the thermal model prediction. 
The discrepancy at RHIC energy, if experimentally
established, would point to a new production mechanism not contained in the
thermal approach and not present at lower beam energies. 
The possible existence of an excited $J^\pi=3/2^+$ state of (anti)hyper-tritium
has been recently pointed out to us \cite{a_gal}.
This excited state could contribute via decay to the ground state, and would lead to
close agreement between model and data.
Further measurements at RHIC and, very soon, LHC energy are eagerly awaited 
to shed light on the situation.

\begin{figure}[hbt]
\centering\includegraphics[width=.6\textwidth]{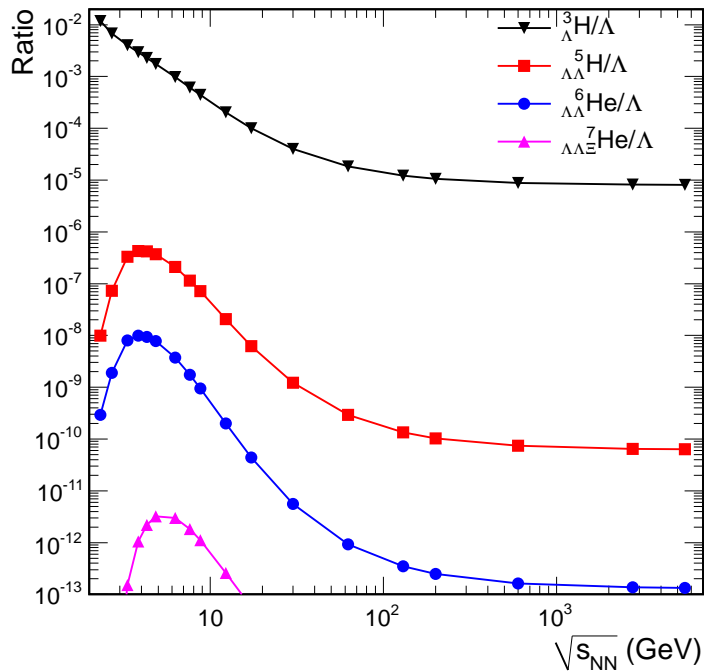}
\caption{Energy dependence of hypernuclei to $\Lambda$ yield ratios.}
\label{fig5}
\end{figure}

To complete our studies we show, in Fig.~\ref{fig5},  predictions within the 
thermal model for the energy dependence of the production yield of multistrange 
light hypernuclei \cite{pbm95} relative to $\Lambda$ hyperons.
These ratios exhibit a pronounced maximum in the FAIR energy regime, 
which is the consequence of a competition between a strong increase 
(followed by saturation) of $T$ and a strongly decreasing $\mu_b$ (see also
the discussion above). 
In addition, the canonical suppression, arising from the condition of local
strangeness conservation, leads to reduced yields at low energies.
In case of hyper-tritium production, there is no maximum, since it is mainly
determined by the strong energy dependence of $\mu_b$ at low energies.
It is larger at the (low) FAIR energies by two to three orders of magnitude 
compared to RHIC and LHC energies. Even larger are the differences between low 
and high energies for the production of the exotic multi-hyperon states.

\begin{figure}[hbt]
\centering\includegraphics[width=.65\textwidth]{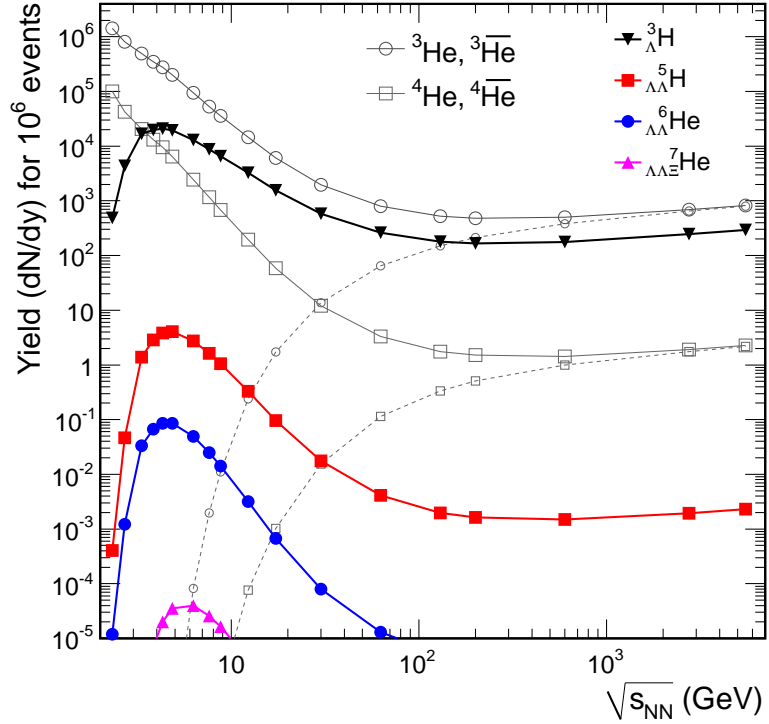}
\caption{Energy dependence of predicted hypernuclei yields at midrapidity
for 10$^6$ central collisions. The predicted yields of $^{3}He$ and $^{4}He$
nuclei are included for comparison, along with the corresponding anti-nuclei 
(dashed lines).}
\label{fig6}
\end{figure}

In Fig.~\ref{fig6} we show, as a function of energy, predictions of yields 
at midrapidity per one million central collisions.
The volume at chemical freeze-out is that from our fits of yields \cite{aa09}.
At FAIR energy the production yields of exotic nuclei is maximal, although
the absolute yields are still rather small.
As an example, for $^7_{\Lambda\Lambda\Xi}He$, the rate of production
at 10$^6$ central Pb+Pb collisions per second is about 60 per month, for 
a reasonable duty factor of the accelerator. Assuming a reconstruction 
efficiency of the order of a percent, this implies a few candidates 
per year of data taking, clearly at the edge of achievability.

At the LHC, (anti-)$^4$He and their corresponding hypernuclei are experimentally 
accessible.
For the LHC energy of 2.76 TeV of the present data taking, we predict ratios 
$^3$He/$^4$He and $^3\bar{\mathrm{He}}$/$^4\bar{\mathrm{He}}$ of 2.76$\cdot 10^{-3}$
and 2.70$\cdot 10^{-3}$, respectively (to be compared to the corresponding values 
for the RHIC energy of 200 GeV of 3.13$\cdot 10^{-3}$ and 2.37$\cdot 10^{-3}$).
These predictions can also be used as guideline for expectation in pp collisions
at  LHC energy, where one could estimate, in the grand-canonical limit, yields
reduced by a factor of the order of 200-400 compared to Pb+Pb collisions.
This is compensated by the much larger number of pp collisions (about 10$^9$ 
events) which can be inspected at LHC (for a running time of 10$^7$ s per year 
of operation).
This should allow, at the LHC, a measurement of the yields of produced
(anti-)hypernuclei up to mass number 4 in pp and Pb-Pb collisions and provide
a detailed test of our predictions.

\section{Conclusions}
We have demonstrated that the yield of light nuclei and their anti-particles
are well reproduced with thermal model calculations employing parameters
established from the analysis of general hadron production in relativistic
nuclear collisions. As shown above, such ratios can be used to provide a
precision constraint of the baryo-chemical potential $\mu_b$. We have
furthermore shown that the newly measured yield ratio
$^{3}_{\bar{\Lambda}}\bar{H}/^{3}_{\Lambda}H$ is also well described with the
thermal approach, while the ratio $^{3}_{\Lambda}H/^{3}He$ which is reproduced
at AGS energy is significantly
underpredicted at RHIC energy. The origin of this discrepancy is currently 
not clear and needs further study. 

Our studies have also indicated interesting energy dependence in such yields
and ratios. In particular, particles with large baryon number and moderate
strangeness are produced in significant numbers at FAIR energy.

The hyper-nuclei program, started by the STAR experiment at RHIC, has made these 
studies very topical. Although significant questions remain, it is clear that 
the study of the production of complex nuclei with and without strangeness in 
relativistic nuclear collisions can open a new chapter in the quest to understand 
the relation of particle production to the QCD phase boundary.
The thermal model predictions can hopefully soon be tested also at the LHC energy
with the data already collected in 2010.

\end{document}